\documentclass[aps,pre,twocolumn,groupedaddress,showpacs]{revtex4}
\usepackage{amsmath}
\usepackage{graphicx,color}
\definecolor{r}{rgb}{1,0,0}
\definecolor{g}{rgb}{0,1,0}
\definecolor{b}{rgb}{0,0,1}

\begin{document}


\title{A Simple Analytical Model of Evaporation in the Presence of Roots}



\author{Cesare M. Cejas$^{1}$, Larry Hough$^{1}$, Jean-Christophe Castaing$^{2}$,
Christian Fr$\acute{e}$tigny$^{3}$, and R$\acute{e}$mi Dreyfus$^{1}$}
\affiliation{$^{1}$Complex Assemblies of Soft Matter, CNRS-Solvay-UPenn UMI 3254, Bristol, PA 19007-3624, USA}
\affiliation{$^{2}$Solvay Research and Innovation Centre - Aubervilliers, France 93300}
\affiliation{$^{3}$Physico-chimie des Polym$\grave{e}$res et des Milieux Dispers$\acute{e}$s CNRS PPMD UMR 7615 ESPCI, Paris, France 75005}


\date{\today}

\begin{abstract}
Root systems can influence the dynamics of evapotranspiration of water out of a porous medium.  The coupling of evapotranspiration remains a key aspect affecting overall root behavior. Predicting the evapotranspiration curve in the presence of roots helps keep track of the amount of water that remains in the porous medium. Using a controlled visual set-up of a 2D model soil system consisting of monodisperse glass beads, we first perform experiments on actual roots grown in partially saturated systems under different relative humidity conditions. We record parameters such as the total mass loss in the medium and the resulting position of the receding fronts and use these experimental results to develop a simple analytical model that predicts the position of the evaporating front as a function of time as well as the total amount of water that is lost from the medium due to the combined effects of evaporation and transpiration. The model is based on fundamental principles of evaporation flux and includes empirical assumptions on the quantity of stoma in the leaves and the transition time between regime 1 and regime 2. The model also underscores the importance of a much prolonged root life as long as the root is exposed to a partially saturated zone composed of a mixture of air and water.  Comparison between the model and experimental results shows good prediction of the position of the evaporating front as well as the total mass loss from evapotranspiration in the presence of real root systems. These results provide additional understanding of both complex evaporation phenomenon and its influence on root mechanisms.
\end{abstract}

\pacs{to be determined}
%

\maketitle




Evapotranspiration from a porous medium, a combined process of evaporation and transpiration, is an important water transport mechanism involved in soils containing living organisms such as plants. The diversity of the potential applications in which evapotranspiration plays an important role makes it a widely investigated case study of scientific and technological relevance. During the phenomenon of evapotranspiration, water uptake occurs within through two forms: on one hand, water is evaporated at the surface of the soil while on the other hand, water is absorbed by the roots of the plants in the soil and is subsequently evaporated at the surface of the leaves.

The former form, water absorption by roots, has been modelled extensively. The earliest model on water uptake was developed by Gardner~\cite{Gardner60}. It describes the extraction of water contained in a cylinder of soil around a root. Though simplistic, its main contribution is analytically showing a decrease in water content in the vicinity of the root. It has been shown that fluctuating water content affects root activity~\cite{Bengough11, Whalley05a, Cassab13}. Since then, succeeding models on water uptake have leaned towards simulations~\cite{Doussan06, Clausnitzer94, Schroder08} to physically model experimental results, introducing different factors due to the complex nature of the root activity. These factors often couple water extraction and root growth. One example is the correlation of water content in the soil profile with root architecture~\cite{Doussan06, Garrigues06}, showing decreasing water saturation as a function of distance away from the root and also as a function of the root system topology, whether fibrous or taprooted. 

The latter form, evaporation from a porous medium, has also been considerably studied. Several experimental and theoretical studies~\cite{Lehmann08, Shokri09} have described the challenges of predicting the evaporation behavior and the drying rates owing to the different mechanisms that are involved at the pore scale level, such as pore impregnation and drying~\cite{Prat07}, liquid flows due to pore size distribution~\cite{Shokri09}, and corner flows due to grain contacts in the porous medium~\cite{Yiotis12}.

Although water evaporation and water absorption by roots have been separately studied, the aspect of evapotranspiration, where both process occurs and are coupled, has not been widely investigated due to the inherent complexity of the fluxes involved~\cite{Pierret07, Bengough11}. This research could re-invigorate important insights into water dynamics in the porous medium in the presence of roots and in the face of on-going evaporation.

We therefore present here an analytical model for evapotranspiration in the presence of roots. This model is developed based on experimental results of the response of root systems under controlled evaporation conditions. Using a 2D visual experimental set-up of root growth in a model porous medium, the model shows the contribution of both evaporation and transpiration fluxes to overall water uptake in the porous medium that results from varied evaporation conditions. This simple analytical model of evapotranspiration with real root systems takes into account the importance of root exposure to an unsaturated or partially saturated zone as a major factor in keeping the root alive and preventing it from dying. These results of the model are able to predict the amount of water that has evaporated and transpired from the granular medium in the presence of roots.

\section{Experimental Set-up}

We perform root growth experiments in a two-dimensional (2D) set-up, also known as rhizotrons~\cite{Futsaether02, Garrigues06}, using a pair of glass sheets put together to make a Hele-Shaw cell of size $10$~cm x $15$~cm. All of the sides are sealed with a commercial silicon paste and only the upper portion remains open permitting evaporation when the cell is oriented in a vertical manner. Because of the silicon paste, this effectively reduces the size of the porous medium to $7.5$~cm x $13.5$~cm. The cells are filled with monolayer glass beads of diameter, $d = 1 \pm 0.2$mm (borosilicate, Sigma-Aldrich), thus the thickness, $e$, of the 2D growth cell is roughly the diameter of the beads. The glass sheets are made hydrophobic to reduce wetting effects using a hydrophobic silane solution (OMS Chemicals), while the glass beads, however, are hydrophilic for all root elongation experiments. Since glass is naturally hydrophilic, they are simply washed with $0.1$~M HCl and dried in an oven overnight at $70^{\circ}$C. Contact angle values are $82 \pm 4^{\circ}$ for hydrophobic glass sheet and $16 \pm 2^{\circ}$ for hydrophilic glass beads. 

Studies on roots often pose a challenge because they normally grow in opaque environments~\cite{DeSmet12} and thus research often relies on the use of artificial conditions. However, $2$D environments are common due to their relative simplicity in terms of visualization~\cite{Clausnitzer94, Futsaether02, Doussan06, Garrigues06, Cejas13b} as well as their reproducibility. We select lentil roots, $Lens$ $culinaris$, as our root because they grow easily and their overall root system is relatively simple~\cite{Futsaether02}. In addition, the size of the lentil root is big enough to be easily distinguished using optical tools and to be easily contrasted with the front patterns that develop from evaporation. 

A sample image of the $2$D root growth cell is shown in Fig. $1$, where a lentil root is grown inside the cell.

\begin{figure}
\includegraphics[width=3in]{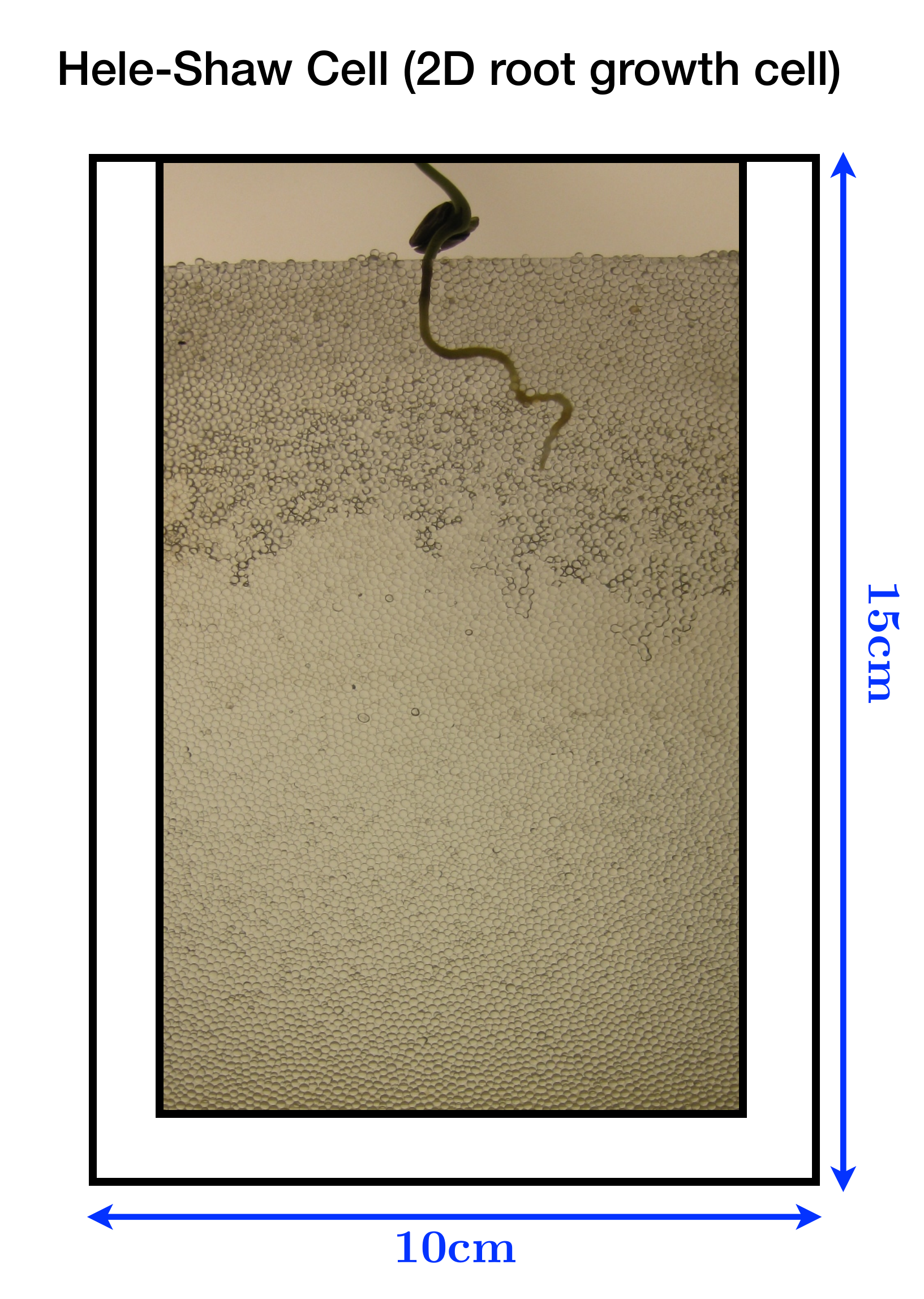}
\caption{(Color online) A sample image of a lentil root grown inside a 2D Hele-Shaw cell of width 10~cm and length 15~cm. Due to the silicon paste present on the sides, the effective length is reduced to 7.5~cm x 13.5~cm. }
 \label{Fig1}
\end{figure}

The 2D cells filled with glass beads are then fully saturated with water mixed with an amount of Hoagland nutrient solution ($25\%$ w/v). The nutrients are especially essential during the initial growth phase of the roots. We use a vacuum pump to saturate the pore spaces of the medium by fully immersing the cell in a bath filled with the liquid. The vacuum displaces the air in the pore spaces with water and helps remove unwanted bubbles.

Lentil seeds are initially germinated for $2-3$ days in the dark on a moist tissue. Then, once a radicle appears, we carefully select the seeds to ensure that at the beginning of our experiments they all roughly have the same mass before they are placed on the $2$D cell. The $2$D growth cells are placed beneath a white lamp as grow light. The average length of radicle after germination is around $L_T = 1.5\pm0.5$cm. The roots are then grown under ambient conditions at temperature $T = 23\pm2^{\circ}$C and relative humidity $R_H = 45\pm5\%$ to foster initial root growth. Immediately exposing the radicles to high evaporation demand does not result to growth. It is also reasonable to expect that the initial phase of root growth takes considerable time since the root is still developing its early biological functions. For this reason, during the first few days following transplant, normally $2-3$ days, the radicles that have just sprouted from the seed and transferred unto the 2D cell are first grown in semi-fully saturated conditions to prevent quick death. This means that as water evaporates, it is being replenished. After which, the cell is ready for evapotranspiration experiments.

Once the radicle has sufficiently developed, the $2$D cell is transferred to a controlled environment chamber (Model $5100$ Electro-Tech Systems) to start evaporation experiments under controlled conditions and at high evaporative demand, particularly at a different temperature and relative humidity, $T = 32\pm2^{\circ}$C and $R_H = 20\pm2\%$. The controlled chambers have dual humidification and dehumidification systems allowing control of temperature and relative humidity. The 2D cells are placed on top of a balance to record mass loss as a function of time and images of the root growth and evaporation are taken at specific time intervals using a Canon $500$D SLR camera using a $18-55$mm lens. We use another light source to back-illuminate the cells to effectively contrast the water distribution. The presence of the light source near to the cell helps keep the temperature constant at $T = 32\pm2^{\circ}$C.

In all the experiments under high evaporative demand, the cell always starts out fully saturated. The obvious challenge when studying root systems is that the duration of the experiment usually lasts relatively long depending on the growth characteristics of the chosen plant. Root growth test runs under ambient conditions prior to the actual experiments reveal similar growth patterns over a long duration.

Roots in granular materials normally adopt a curved profile. As a result, the contour length is measured using the segmentation method by creating smaller line segments, whose total length adds up to the total root contour length. Preliminary tests of the segmentation method performed on various sets of curves of known length reveal a measurement error within $\pm 5\%$. The total length is measured as a summation of the lengths of the individual line segments.

\begin{figure*}
\includegraphics[width=6.8in]{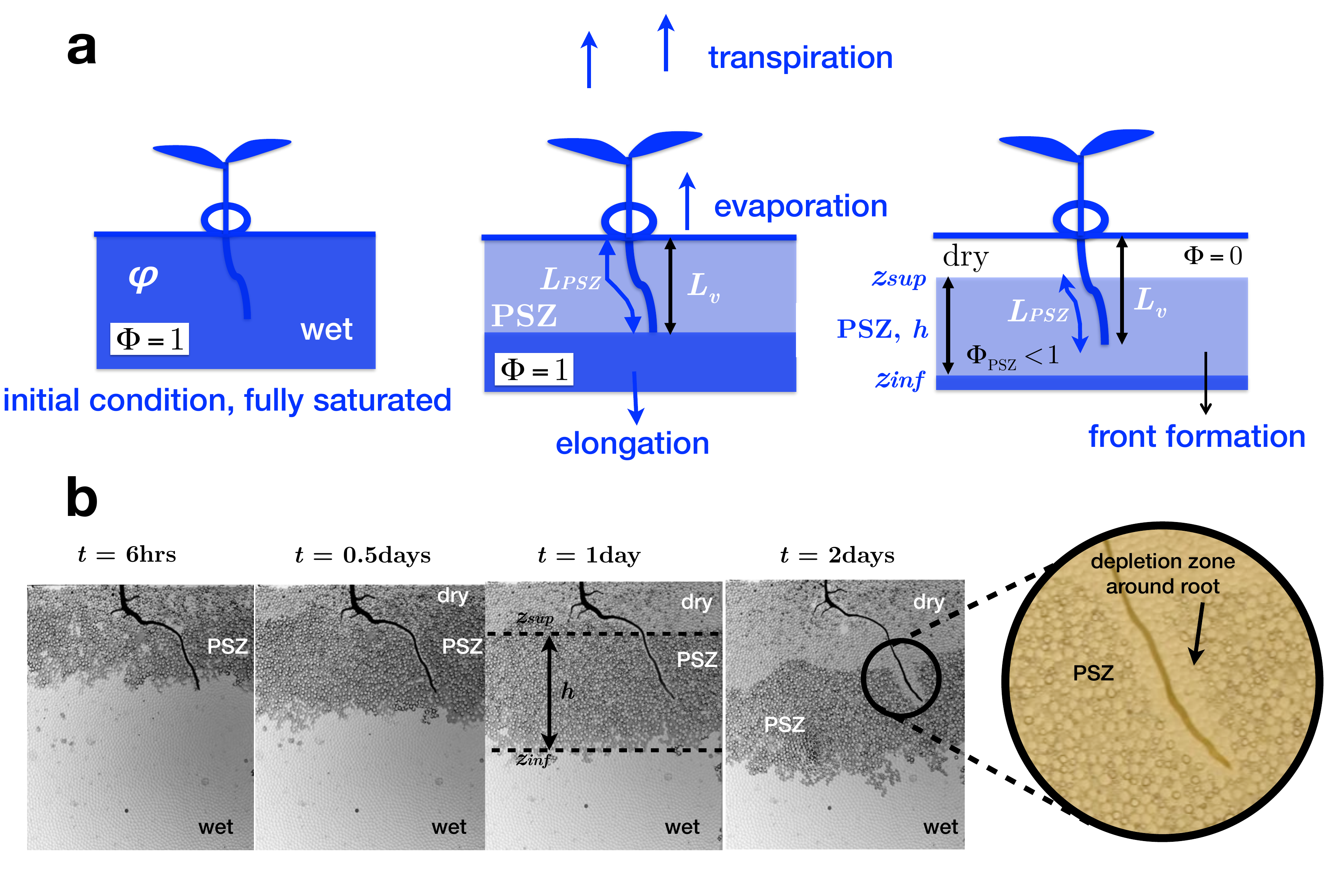}
\caption{(Color online) (a) Illustration of the experiment involving root elongation in a medium of porosity, $\varphi$. The system starts out fully saturated, $\Phi = 1$. The unsaturated system results from the evaporation of water from the medium that leads to the appearance of the partially saturated zone (PSZ), $h = z_{inf} - z_{sup}$, a mixture of air and water ($\Phi < 1$). As water evaporates, a portion of the root is exposed to a partially saturated zone, $L_{PSZ}$, especially during the first regime. The vertical distance between the tip of the root and the cell surface is $L_v$. During the first regime, transpiration rates are therefore high. But during the second regime, a dry region ($\Phi = 0$) develops above the receding evaporating front, $z_{sup}$, which  slows down evaporation. (b) Experimental images showing the evolution of the front during evaporation in the presence of roots. The roots absorb water to replenish the amount of water lost and is clearly manifested by the appearance of a depletion zone around the roots. }
 \label{Fig2}
\end{figure*}

\section{Results and Discussion}
\subsection{Characterization of evaporation with roots}

The presence of roots in the granular medium will clearly affect water distribution and thus it is essential to characterize the evaporation dynamics in the presence and absence of root systems. Evaporation with roots is made complicated by an additional flux that has to be taken into account. This flux is transpiration, which is the evaporation of water from leaves~\cite{Nobel09, Larcher95} and results from the opening and closing of stomata cells during their intake of carbon dioxide essential for photosynthetic activity. These stomata cells are found at the surface of plant leaves. 

During this process of gas-exchange, water vapor simply diffuses out of the stomata. The loss of water in the leaves induces a pressure differential across the plant, which results into absorption by the roots to compensate for water loss. Transpiration gives a negative potential of water around the roots. This results to water absorption that also leads to differential cell expansion that helps root elongation. 

An illustration of the evaporation process with roots is presented in Fig. 2b. Fig. 2b shows some experimental images of evaporation of water in a 2D cell in the presence of roots. the roots are initially exposed to a fully wet zone, $\Phi = 1$. The evaporation process creates a partially saturated zone (PSZ), $\Phi < 1$. Over time, a dry zone, $\Phi = 0$, eventually catches up with the root until the entire root system is exposed to it.  A clear manifestation of the absorption process and thus the root activity is seen from the occurrence of the depletion zone around the root. Under high evaporative demand, root growth inside the 2D model porous medium can be extremely minimal and root growth rates are extremely slow compared to the rate of formation of the evaporating front. Therefore, root elongation under these conditions can be considered almost negligible as supported also by the minimal change in root length values over time. 

\begin{figure}
\includegraphics[width=3.5in]{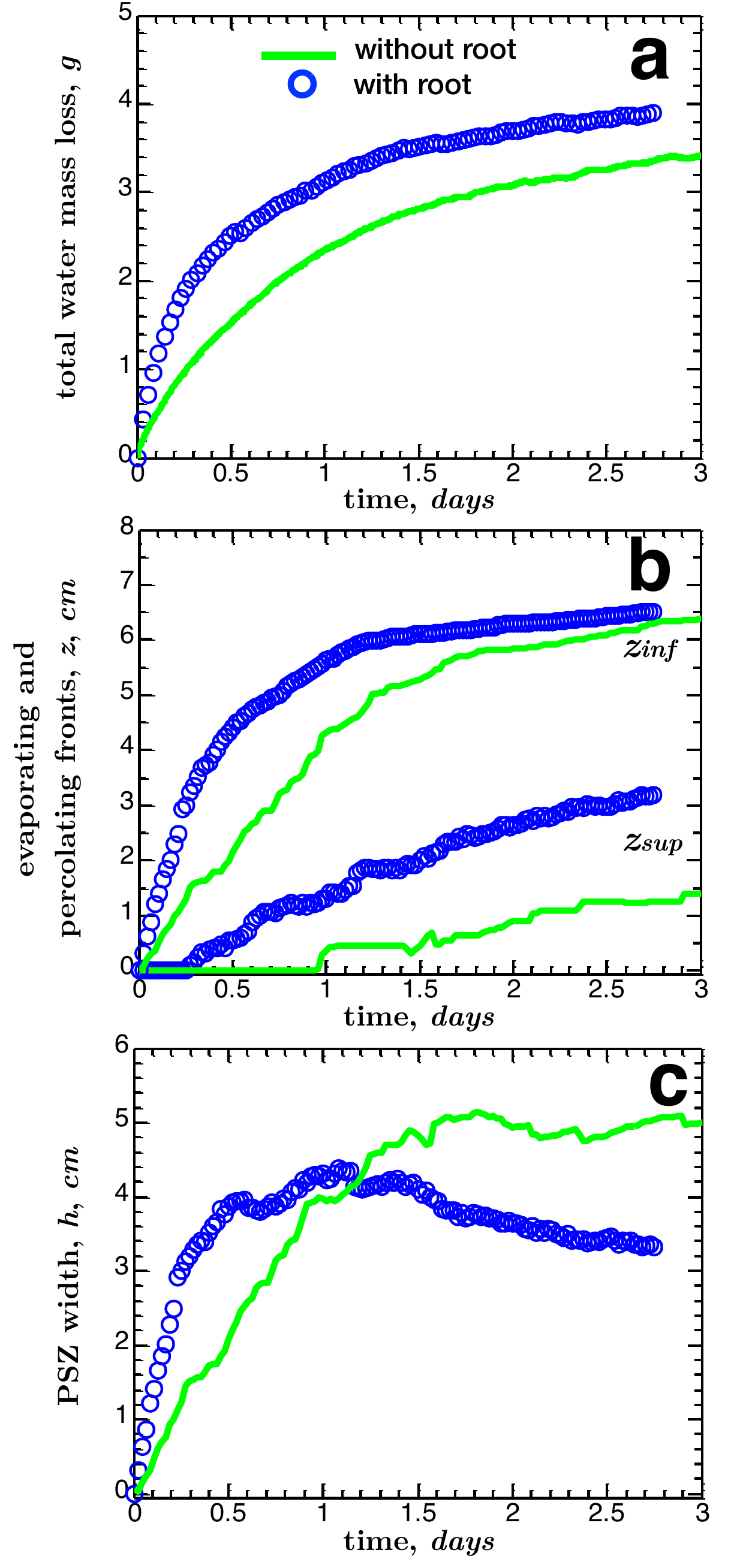}
\caption{(Color online) Comparison of evaporation dynamics of water in 2D Hele-Shaw cell in the presence and absence of root systems. (a) Mass curves depicting total mass loss in the root and granular system as a function of time. (b) Measured evaporating and percolating fronts denoted as $z_{sup}$ and $z_{inf}$ respectively. (c) Measured height of the PSZ, $h = z_{inf} - z_{sup}$. These curves have been obtained under conditions of $T = 32\pm2^{\circ}$C and $H_R = 20.0\pm2.0\%$.   }
 \label{Fig3}
\end{figure}

Generally, we can expect two competing rates. First is the characteristic root elongation rate, directed downwards and second is the evapotranspiration rate directed upwards. The latter can be divided into the transpiration and evaporation components. It is clear that the root elongation rate would ultimately depend on the evaporation conditions as water serves as a major limiting factor in its growth. At low humidity conditions, we expect the evapotranspiration rate to be faster, thus the development of the evaporation front, $z_{sup}$ also occurs fast. When this happens, the evaporating front, $z_{sup}$, eventually catches up with the root tip and quickly exposes a large portion of the total root system to a dry zone. The appearance of the dry zone certainly decreases water absorption rates. 

We characterize the evaporation with and without root systems by measuring total mass loss, positions of the evaporating and percolating fronts, $z_{sup}$ and $z_{inf}$ respectively, and the average size, height or thickness of the PSZ, $h$. 

We plot the evaporation curve in a 2D model system performed under the same conditions in Fig. 3a. At first glance, Fig. 3a shows that evaporation in the presence of roots loses more mass of water compared to the model system without roots. This suggests that the roots provide an additional pathway for water loss. Note that the mass of the plant has not yet been subtracted from this curve. When we take into account the mass of the plant, $m_p$, which is also a function of time, the actual water loss from evaporation with roots is still more pronounced since lentils are very light plants and within the duration of the experiment, there is no significant variation in plant mass.

Experimentally, the evaporating and percolating fronts in the presence of roots show a steep slope in Fig. 3b. This suggests rapid water loss. In addition, the $z_{inf}$ values seem to approach the same final value and is independent of the presence of roots. The presence of the root system, however, clearly affects the formation of the evaporating front, $z_{sup}$. This results to a clear difference in $h$ values as shown in Fig. 3c. Given that all other physical parameters (particle size, cell size, humidity, etc.) are kept constant for both types of experiments, then the difference in $h$ can clearly be attributed to the presence of roots. In Fig. 3c, the value of $h$ in the presence of roots is smaller than without roots. This is an interesting result as it provides a clue into the activity of the root within the PSZ. When more water is lost in the porous medium, this leads to faster formation of the receding fronts yet ultimately leads to a lesser size of the PSZ as seen from these results.

The presence of the roots clearly modifies the total evaporation dynamics of water out of a given porous medium. The next section describes a simple anlaytical model that predicts the total evaporative flux emanating from the porous medium in the presence of a root when the root is exposed to a PSZ and consequently the death of the root when the root is completely exposed to a dry zone. 

\section{Theoretical Considerations}

As can be seen from the experimental observations, two regimes of evapotranspiration are observed. There is a fast regime or regime 1, in which a connection of water exists between the percolating front and the top of the cell~\cite{Prat99, Lehmann08, Shokri09, Yiotis12}. The second regime termed regime 2 appears when this connection is broken at the surface and a dry zone starts to appear just below the cell surface. As water is simultaneously removed from the soil through evaporation and transpiration we can define four different fluxes. $J_{e_1}$ and $J_{e_2}$ corresponds to the flux of evaporation during regime 1 and 2 respectively while $J_{t_1}$ and $J_{t_2}$ corresponds to the flux of transpiration during regime 1 and 2 respectively. If $m$ corresponds to the total mass loss of water, then the rate at which the mass of water is taken out of the soil for both regimes is:

\begin{equation}
   \frac{dm}{dt} = J_{e_{1,2}}S+J_{t_{1,2}}S,
\label{Eq1}
\end{equation}

where $S$ is the cross-sectional area of the surface. We use a plus sign here as we choose to consider as a convention the mass loss of water as a positive value when water escapes from the soil.

In the following subsections, we derive the expressions for these four different fluxes based on parameters that we can measure. These analytical expressions include fitting parameters that are nevertheless constant regardless of relative humidity value. We first start by defining the evaporation flux and tthe transpiration flux.

\subsection{Evaporation flux}

Prior also to performing evaporation experiments in the presence of roots, we have extensively performed experiments involving evaporation of water from the same granular medium but without roots. The first regime of evaporation is described as having high rates due to interconnected air-liquid interfaces that link the percolating front to the evaporating surface. This permits rapid flow of water out of the granular medium. During regime 1, liquid transport out of the cell proceeds via diffusion across an external transfer zone, $\delta_{cell}$, as can be described using Fick$^{\prime}$s law:

\begin{equation}
   J_{e_1} = D_a \left(\frac{C_{sat} - C_e}{\delta_{cell}}\right),
\label{Eq2}
\end{equation}
where $D_a$ is the diffusion of water in the air, $C_{sat}$ is the concentration of water at the evaporation plane, and $C_e$ is the concentration of water in the surrounding atmosphere, which is directly related to the relative humidity, $H_R$. The value of $C_{sat}$ can be calculated from the saturation vapor pressure of water at a particular temperature using the ideal gas law. 

The second regime of evaporation starts when the liquid film connections detach from the surface. As a result, an evaporating front occurs that recedes inside the medium at the same rate as the percolating front. A dry region also develops. Evaporation rates slow down as water now diffuses across this dry region. Similarly, the flux during the second regime $J_{e_2}$ can also be described via Fick$^{\prime}$s law. Note that there are now two gradients building up in the system, a flux through the porous medium and another flux across the external mass transfer zone. 

\begin{equation}
   J_{e_2} =  D_p \left(\frac{C_{sat} - C_s}{z_{sup}}\right),
\label{Eq3}
\end{equation}
where $D_p$ is the diffusion of water across the porous medium, which is equal to $D_a$$\varphi^{1.5}$~\cite{Marshall59, Moldrup00}, $C_s$ is the concentration of water the at cell surface, the value of which is currently unknown. As the value of $C_s$ is unknown, it is important to notice that the flux of water from the interior of the cell to the surface as well as the flux of water from the cell surface to the surrounding atmosphere is derived as:

\begin{equation}
   J_{e_2} =  D_a \left(\frac{C_s - C_e}{\delta_{cell}}\right),
\label{Eq4}
\end{equation}

By equating Eq. 3 and Eq. 4, we can derive an expression for $C_s$. We can use the resulting expression of $C_s$ in any of the equations either Eq. 3 or Eq. 4 and obtain an expression for $J_{e_2}$:

\begin{equation}
   J_{e_2} = \frac{C_{sat} - C_e}{{\frac{z_{sup}}{D_p}}+{{\frac{\delta_{cell}}{D_a}}}},
\label{Eq5}
\end{equation}

Now that we have derived  the equations pertaining to evaporation from a porous medium,  these equations serve as a basis for deriving the equations for evaporation from a porous medium in the presence of root systems.

\subsection{Transpiration flux}

Transpiration occurs in the leaves of plants during photosynthesis when water vapor exits the leaves through the stoma as carbon dioxide is taken in by the plant. The stoma clearly plays a role in gas exchange~\cite{Nobel09, Larcher95}. Plants have an internal regulation that controls the opening and closing of the stoma. The stoma can act like a pump, which pulls water from the roots upward to replenish the quantity of water that had been lost during transpiration. During this process, the stomata open up thereby also increasing transpiration rates. For this reason, we expect the transpired flux to be strong as long as the roots remain in a mixture of air and water or a PSZ. As the dry zone appears and further recedes into the medium, the plant becomes less robust and thus transpiration rates decrease as well.

In this section, we address how the presence of the roots of plants modify the total evaporation dynamic due to the additional transpiration flux. A simple conceptual illustration of the leaf surface is shown in Fig. 4 and helps clarify the transpiration process.


\begin{figure*}
\includegraphics[width=7in]{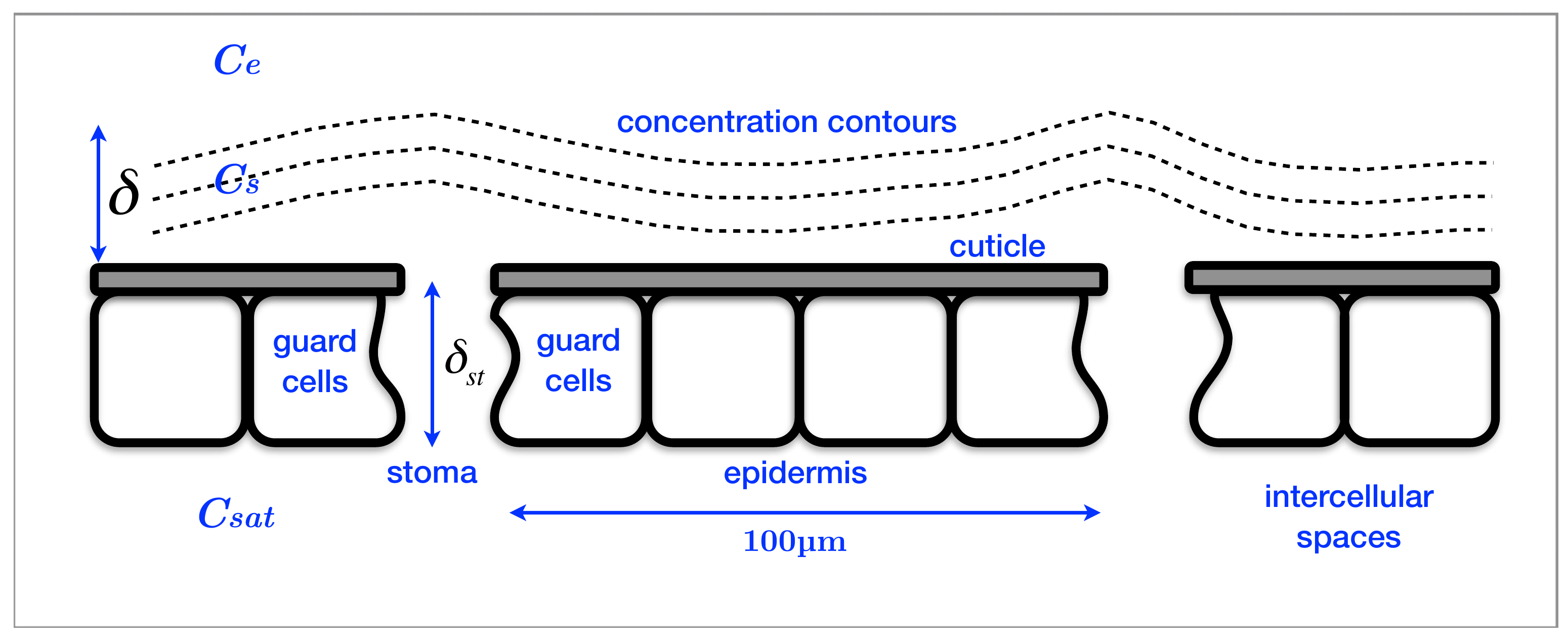}
\caption{(Color online) Illustration of the stoma in the leaf surface (adapted from~\cite{Nobel09}). Water vapor diffuses through the stoma of height $\delta_{st}$ and across a external mass transfer distance of $\delta$. The parameter $C_{sat}$ is the concentration of water inside the leaves, the parameter $C_s$ refers to the concentration of water at the surface of the leaves, while $C_e$ is the water concentration in the surrounding atmosphere. The paremeter $C_e$ is dependent on the relative humidity value.}
 \label{Fig4}
\end{figure*}


We define the way water escapes out of the 2D medium by first imposing the condition that water must diffuse through the stoma in its vapor phase. Water travels from the interior of the leaves to the exterior across a distance $\delta_{st}$, which is the characteristic height of the stomata. Total flux density depends on the size of the surface so the total surface considered here is $S_{st}$, which is the total surface of the stomata. If the concentration of water inside the leaves is $C_{sat}$ and the concentration of water right at the leaf surface is $C_s$, then by Fick$^{\prime}$s law, the total mass of transpired water $m_t$ per unit of time is:

\begin{equation}
   \frac{dm_t}{dt}= D_a \left(\frac{C_{sat} - C_s}{\delta_{st}} \right) S_{st},
\label{Eq6}
\end{equation}

Eq. 6 is the same principle as Eq. 2. All of the water diffusing through the pores also diffuses to the surrounding atmosphere of water concentration $C_e$. Thus, loss of mass per unit of times becomes:

\begin{equation}
   \frac{dm_t}{dt}=  D_a \left(\frac{C_s - C_e}{\delta} \right)S_{L},
\label{Eq7}
\end{equation}
where $S_L$ is the total surface area of the leaf and $\delta$ is the height of the external mass transfer zone at the leaf surface. The parameter $C_s$ is again unknown but by equating Eq. 6 and Eq. 7, we can solve for $C_s$ and re-inject its value in either equation to derive the expression for the transpiration:

\begin{equation}
   \frac{dm_t}{dt} =  D_a\left(\frac{{\Delta}{C}}{{\delta_{st}}+{\delta \left({\frac{S_{st}}{S_L}}\right)}}\right)S_{st},
\label{Eq8}
\end{equation}
where ${\Delta}{C} = C_{sat} - C_e$. By introducing the parameter $\phi_{st}$, which is the fraction of leaf area occupied by the stomata  $ \phi_{st} = S_{st} / S_L $,  we can then turn Eq. 8 to Eq. 9. 
\begin{equation}
   \frac{dm_t}{dt} =  D_a\left(\frac{{\Delta}{C}}{{\delta_{st}}+ { \phi_{st} \delta}}\right) \phi_{st} S_{L},
\label{Eq9}
\end{equation}

\begin{figure}
\includegraphics[width=3.2in]{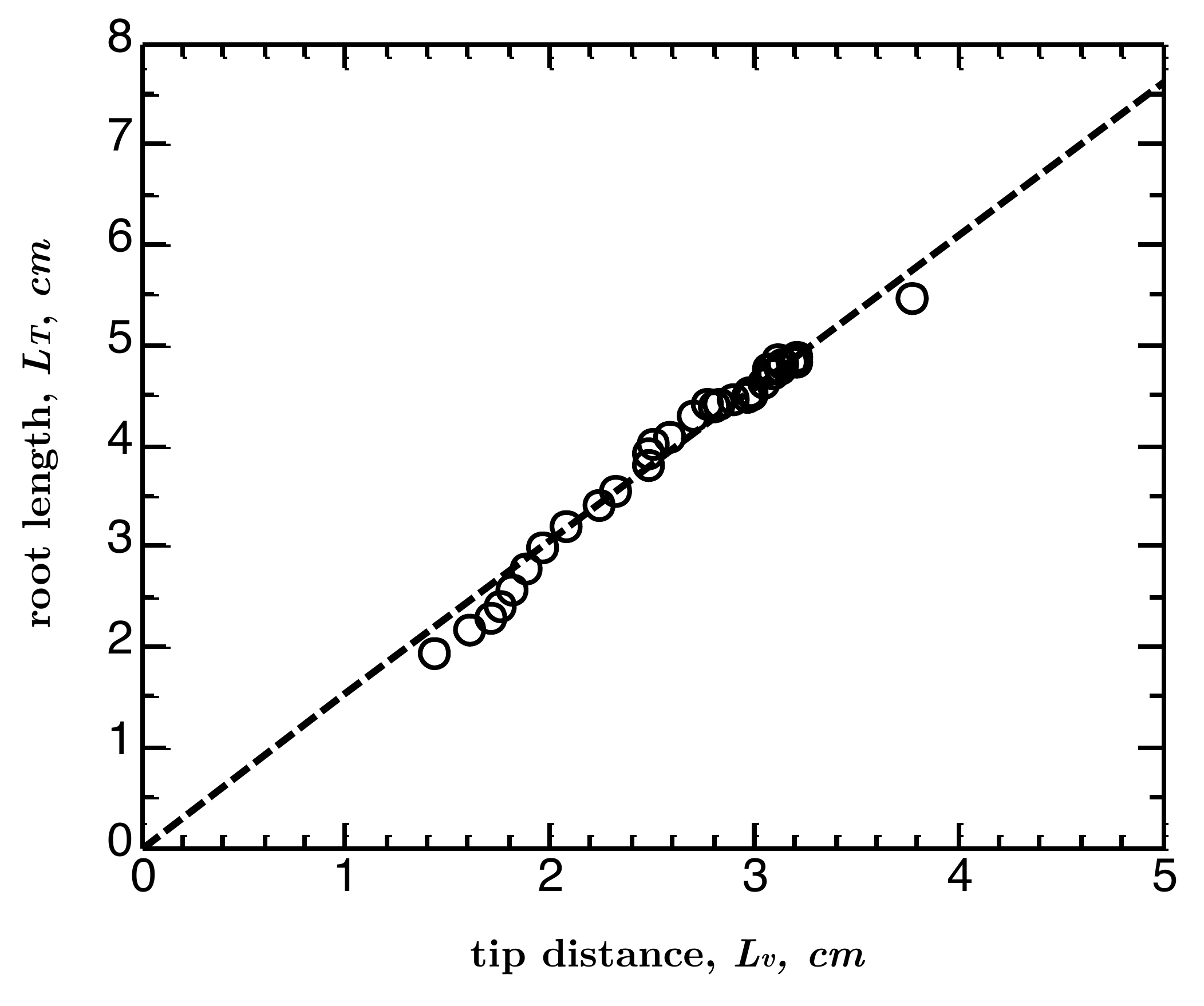}
\caption{(a) Plot of root length, $L_T$, as a function of distance between the root tip and cell surface, $L_v$. These results reflect an average of the experiments performed for the characterization of lentil roots under fully saturated conditions. Experiments show a linear proportionality between root length and tip distance, as seen from the broken line which serves as a guide. }
 \label{Fig5}
\end{figure}

It is unclear how the area of the stomata, $\phi_{st}$, varies. It is evident however that the stoma have a response to external stimuli such as the quantity of water in the medium. So a dependence of $\phi_{st}$ on water content is logical. 

For the sake of simplicity, we assume the following relations in the development of this analytical model.

First, during the regime 1 evaporation when the evaporating front has not yet formed, thus $z_{sup} = 0$ and the root is immersed in a partially saturated zone or PSZ. In this regime, since plants benefit greatly from a PSZ, we assume that the stomata are open, therefore $\phi_{st}$ remain constant.

Second, during regime 2 evaporation when the evaporating front has now formed and receded into the medium, thus $z_{sup} > 0$ and a part of the root system is now exposed to a dry zone, while another part is still immersed in a PSZ, $L_{PSZ}$. The appearance of the dry zone results to water stress that reduces transpiration activity, which can be induced by the closing of the stomata. The simplest assumption that is feasible for the moment is that the surface of the stomata varies linearly with the height of the remaining PSZ in the medium that is exposed to the root, $\phi_{st} \sim \beta$ ($L_v - z_{sup}$), where $L_v$ is the vertical distance between the root tip and the topmost surface of the cell (see Fig. 2a) and $\beta$ is a constant. Experimental results particularly on the extensive characterization of lentil roots in 2D model granular medium have shown a linear proportionality between the parameter $L_v$ and the root length $L_T$ as shown in Fig. 5.

Third, in Eq. 8 it is actually known that when the stoma are open, $\phi_{st}$ varies between $0.1\%$ and $1\%$ depending on plants ~\cite{Nobel09}. Since $\delta \sim 1$~mm, $\delta \phi_{st}$ remains small compared to the height of the stomata, $\delta_{st}$. Over time, this quantity slightly decreases. Therefore, we currently neglect this variation in this model for simplicity.

Fourth and last, we extend the second assumption by considering that there is also some relationship between the surface area of the leaf and size of the root. Again, the simplest relationship is a linear proportionality, $S_L = \alpha L_v$ where $\alpha \sim 1.12.10^{-2}$~m is measured from experiments. 

With these assumptions, we can now obtain the expression for the transpiration flux in both regimes:

During regime 1:

\begin{equation}
   J_{t_1} = D_a \frac{{\Delta}{C}}{{{\delta_{st}}+{\phi_{st}}\delta}} \frac{\alpha \phi_{st} L_v}{S},
\label{Eq10}
\end{equation}
while during regime 2:

\begin{equation}
   J_{t_2}(t) = D_a \frac{{\Delta}{C}}{{{\delta_{st}}+\phi_{st}\delta}} \frac{\alpha  \beta \left( L_v - z_{sup} \right) L_v}{S},
  \label{Eq11}
\end{equation}

As we saw before, the rate of mass of water extracted out of the soil  in the first regime, is related to those different fluxes by this relation:

\begin{equation}
   \frac{dm}{dt} = J_{e_1}S+J_{t_1}S,
\label{Eq12}
\end{equation}

For the first regime, both the evaporation flux and the transpiration flux do not depend on time. Therefore the equation for the first regime can be readily integrated assuming that $m(t = 0) = 0$. We obtain the following expression for the total mass of water that is extracted from the soil during regime 1:

\begin{equation}
   m(t)= (J_{e_1}S+J_{t_1}S)t,
\label{Eq13}
\end{equation}

Eq. 13 predicts a linear relation of mass with time during regime 1, which is consistent with experimental observations (see Fig. 3 and Fig. 6b).

For the second regime:

\begin{equation}
   \frac{dm}{dt} (t) = J_{e_2}(t)S+J_{t_2}(t)S,
\label{Eq14}
\end{equation}

In this regime, integrating the equation is not straightforward as both the evaporated and the transpired flux depend on $z_{sup}$ which is a function of time. At a certain time $\it{t}$ during the experiment, the mass that has evaporated is the sum of the mass of water that was initially contained in the dry zone $z_{sup} \varphi S \rho$ and the mass of water out of the partially saturated zone $h S \rho \varphi (1-\Phi)$.

\begin{equation}
 m(t) = -{z_{sup}(t) \varphi S \rho} + {h S \rho \varphi (1-\Phi)},
\label{Eq15}
\end{equation}

By injecting Eq. 15 to solve Eq. 14, we obtain a differential equation for $z_{sup}$. We assume that there is a certain critical evaporated mass, $m_c$, above which the system transitions from regime 1 to regime 2. The parameter $m_c$ corresponds to a certain time $t_c$, thus, $m(t_c) = m_c$. Then, we obtain the following expression for $z_{sup}(t)$ in Eq. 16, during the second regime:

\begin{equation}
   t = t_c + \frac{\rho \varphi}{D_p \Delta C(2 \overline{A}^{\ast} F_1)} \Pi,
\label{Eq16}
\end{equation}
where $\Pi$ represents:

\begin{equation}
   \Pi = F_2 \left(\text{arcth}(F_3) - \text{arcth}(F_4)\right)+  F_1 \text{ln}(F_5),
\label{Eq17}
\end{equation}

In Eq. 17, the following variables are represented as follows:

\begin{equation}
\begin{split}
F_1 &=\sqrt{4 + \overline{A}^{\ast}(L_v + \overline{D})^2}, \\
F_2 &= 2 \sqrt{\overline{A}^{\ast}} (L_v + \overline{D}), \\
F_3 &= \frac{\sqrt{\overline{A}^{\ast}} (\overline{D} - L_v + 2 z_{sup})}{F_1}\, \\
F_4 &= {\frac{\sqrt{\overline{A}^{\ast}} (\overline{D} - L_v)}{F_1}}, \\
F_5 &= \frac{1 + \overline{A}^{\ast} (L_v - z_{sup})(\overline{D} + z_{sup})}{1 + \overline{A}^{\ast} L_v \overline{D}},
\label{Eq18}
\end{split}
\end{equation}





The parameter $\overline{D}$ is defined as $\overline{D} = \frac{D_p}{D_a} \delta$ while the parameter $\overline{A}^{\ast}$ is defined as $\overline{A}^{\ast} = A^{\ast} \frac{D_p}{D_a}L_v$, where:

\begin{equation}
A^{\ast} = \frac{\alpha  \beta}{S\left(\delta_{st} + \phi_{st} \delta \right)},
\label{Eq19}
\end{equation}

Eq. 16 is an implicit equation that considers the following initial condition: that $z_{sup}$ during regime 1 up until the transition time, $t_c$, is equivalent to $z_{sup}$($t \rightarrow t_c$) $= 0$. By setting $z_{sup} = L_v$, this means that entire root is already in the dry zone and will likely die soon and thus in effect, we can view Eq. 16 as an analytical expression of the plant$^{\prime}$s lifetime.

\subsection{Application of model to experiments}

We then first apply the model to the evaporation curves obtained in the presence of root systems from experiments performed at $H_R = 20.0\pm2.0\%$ and $T = 32\pm2^{\circ}$C. In these series of experiments, we record the positions of the evaporating front, $z_{sup}$, and the evaporation curve of the total mass loss in the medium. The temperature value inside the glove box remains constant at $T = 32\pm2^{\circ}$. At this temperature, root length is limited and is almost negligible. 

First, we begin with regime 1 where mass loss can be calculated using Eq. 13. In this regime, there is only one fitting parameter that sets the rate: $\phi_{st}$. We find that the value of the best fit gives $\phi_{st} \sim 0.2\%$, which is consistent with what is reported in literature~\cite{Nobel09}. The evaporating front, $z_{sup}$, is then calculated using Eq. 15. 

In determining the transition time between regime 1 and regime 2, we use a fitting parameter, $m_{fit}$, which corresponds to the critical evaporated mass, $m_c$ at the transition time, $t_c$. The value of $m_{fit}$ is $m_{fit} \sim 2$~g, which is a reasonable value based on the experimental mass curves. Normally, without the presence of roots and in glass bead diameter of $1$~mm, the mass loss during regime 1 is approximately $3$~g. Since the transition time occurs much quicker in the presence of roots, it is logical that the amount of water mass that is lost during regime 1 is less than $3$~g. 

Once $m_{fit}$ has been determined, we now shift attention to regime 2. In regime 2, we assume $\phi_{st} \simeq \beta$($L_v - z_{sup}$). Here, $\beta$ is a fitting parameter that defines the proportionality relationship between the quantity of the root still within the PSZ and the surface area of the remaining available stomata. This is because during regime 2, the appearance of a dry region reduces the length of the root exposed to the PSZ. Therefore, the quantity of the root that is still within the PSZ is the limiting factor in maintaining transpiration rates. We find that the best fit gives a value of $\beta \approx 0.1$~m$^{-1}$. Using these values, we then compare the model with the experimental data. The results are shown in Fig. 6. It is remarkable that by keeping all the parameters constant but changing relative humidity to $H_R = 40.0 \pm 2.0\%$ and $H_R = 75.0 \pm 2.0\%$, the model shows true predictive power as can be seen in Fig. 6. 

Fig. 6a shows the position of the evaporating front $z_{sup}$ as a function of time for different root systems grown in three different relative humidity conditions. The solid curves are calculated from the implicit equation in Eq. 16. The model shows considerable agreement with experiments. Since $z_{sup}$ only begins to appear during regime 2 when it recedes inside the medium, the starting point of each curve therefore reflects the transition time between the first and second regimes of evaporation. From the figure, $z_{sup}$ recedes deeper in higher evaporation conditions or at lower $H_R$ values. 

Fig. 6b shows the mass loss curves from evapotranspiration in the presence of roots as a function of time also for three different relative humidity values. The solid curves are calculated from the definitions of mass loss for both regimes. Determining the transition time between regimes 1 and 2 of evaporation still remains to be a topic of discussion in literature~\cite{Shokri11}. But in this case, the transition time, $t_c$, can be calculated using Eq. 13 using the fitting parameter $m_{fit}$.  Application of these equations also shows good agreement between the model and the experimental data. In this figure, lower $H_R$ denotes higher evaporation conditions and thus more mass of water is lost from the medium.

\begin{table}
\caption{Values of parameters used in the model}
\resizebox{8.5cm}{!}{
\begin{tabular}{c c c c}
	\hline
	{\textbf{Parameter}} & {\textbf{Definition}} & {\textbf{Value}} & {\textbf{Unit}} \\
	\hline
	$D_a$ & diffusion of water in air & $2.5$.$10^{-5}$ & m$^2$~s$^{-1}$ \\
	$D_p$ & diffusion of water in porous medium & $D_a \varphi^{1.5}$ & m$^2$~s$^{-1}$ \\
	$S$ & surface area of the cell & $9.8$.$10^{-5}$ & m$^2$ \\
	$\delta$ & external mass transfer length & $1.8$.$10^{-3}$ & m \\
	$\rho$ & density of water & $1000$ & kg m$^{-3}$ \\
	$\varphi$ & porosity of 2D medium & $0.65$ & no unit \\
	\hline
\end{tabular}
}
\end{table}

\begin{figure}
\includegraphics[width=3.4in]{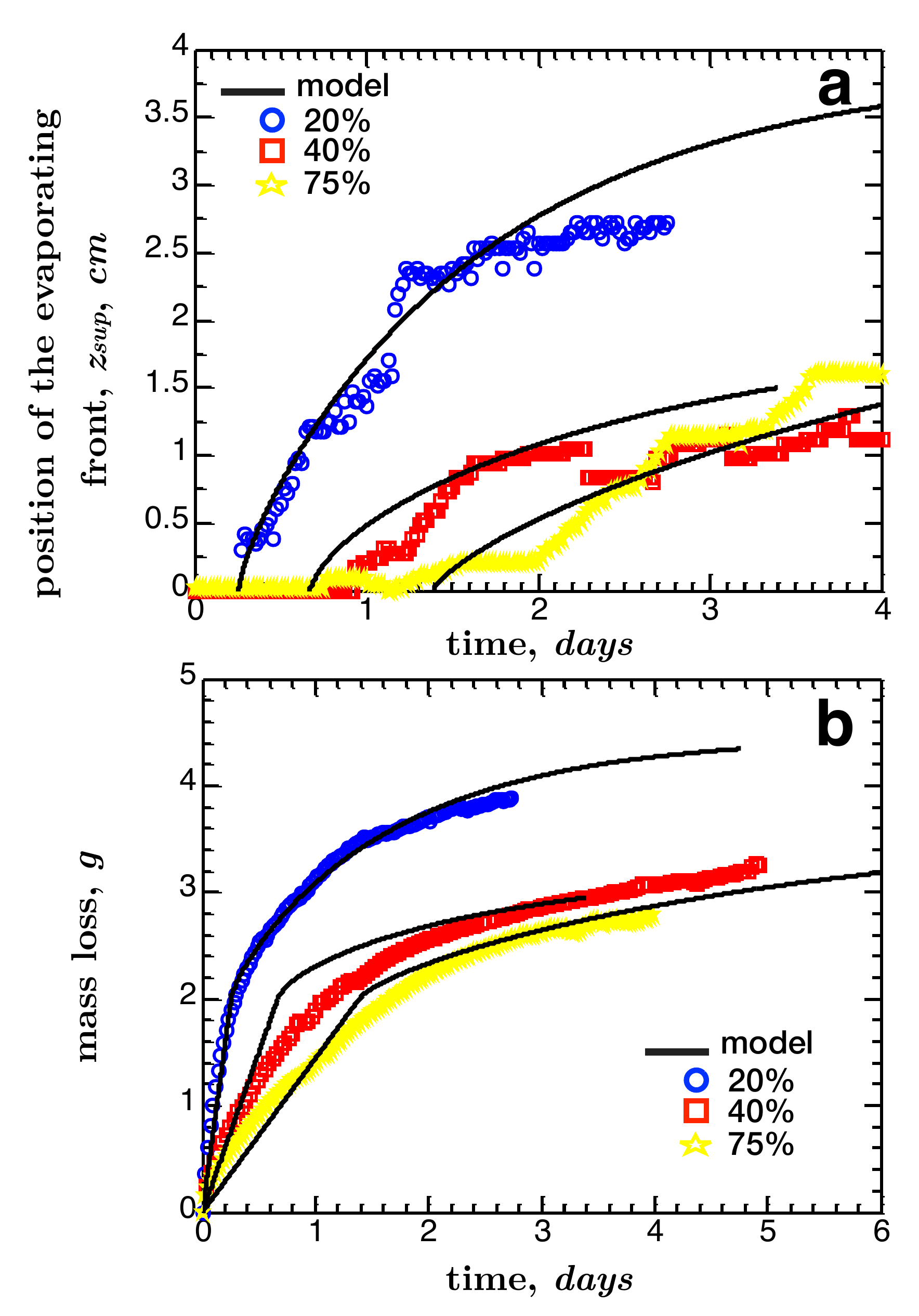}
\caption{(Color online) (a) Position of the evaporating front, $z_{sup}$, as a function of the different roots grown in various relative humidity conditions. The solid curves are calculated from the implicit equation obtained in Eq. 16. (b) Evapotranspiration curves in the presence of roots showing mass loss as a function of time also for different relative humidity conditions. The solid curves are calculated from  Eq. 13 for regime 1 and Eq. 15 for regime 2. }
 \label{Fig6}
\end{figure}

We stress that these fitting parameters have initially been calculated and set for experimental results under $H_R = 20.0 \pm 2.0\%$ but nevertheless remarkably work well for different relative humidity conditions, namely at $H_R = 40.0 \pm 2.0\%$ and $H_R = 75.0 \pm 2.0\%$. Under the given relative humidity and temperature conditions, root elongation is very minimal and slow and thus root length is basically negligible over time. We also assume that despite the use of different lentil plants for each experimental trial at various conditions, the parameters $\delta_{st}$, $\phi_{st}$, $\delta$, and $\alpha$ remain constant.  

Based from the experimental results and fundamental principles of evaporation flux, we have thus established a simple analytical model that estimates and predicts the total evapotranspiration flux in the presence of roots as well as the positions of the resulting evaporating fronts.

\section{Conclusion}

Recent experiments~\cite{Cejas13b} have shown that the presence of a partially saturated zone, PSZ, where $\Phi < 1$, around root systems help increase overall root lengths suggesting that there is a reaction mechanism within roots that is triggered when the roots are exposed to a mixture of air and water instead of a fully saturated zone, where $\Phi = 1$. Using this principle, we have developed an analytical model that predicts the evaporation pattern in the presence of roots in three different external water concentration levels.  Using a simple model soil system, we record the position of the evaporating front, $z_{sup}$ and the total mass loss curves resulting from the combined effect of evaporation and transpiration. We define equations with physical and empirical basis and use fitting parameters which are constant regardless of the value of the relative humidity. 

Though simple and assumptions have been made, this model of evaporation in the presence of roots corroborates results showing that roots grow better in a partially saturated region as opposed to a fully wet one. As along as the roots are in a PSZ, the transpiration rates remain high.  When this is the case, this also implies higher water absorption, which can helps in root cell expansion and helps keep the root alive. The favorable root behavior in the PSZ is a reasonable assumption since too much water may indeed result to root asphyxiation. However, there should also be a maximum PSZ saturation in which root growth is optimal since too little water can also cause root dehydration. These insights are interesting starting points for future investigations and redefinitions of the model. 

Nevertheless, the results presented here still offer fresh insights into the quantification of the water loss during the combined effects of evaporation and transpiration; thereby, improving efforts to estimate plant lifetimes - a key aspect in applications such as water retention.

\begin{acknowledgments}
We thank Solvay, Inc. and the CNRS who have supported this study. We thank everyone in the COMPASS (UMI 3254) laboratory and all those involved and who have contributed to the water retention project. 
\end{acknowledgments}


\bibliography{Roots_References}


%
\end{document}